\documentstyle[11pt,newpasp,twoside]{article}
\def\edcomment#1{\iffalse\marginpar{\raggedright\sl#1\/}\else\relax\fi}
\reversemarginpar

\begin{document}

\title{The Structure of the Circumstellar Envelope of SN 1987A$^1$}

\author{Arlin P.S.~Crotts$^2$}
\affil{Dept.~of Astronomy, Columbia University, New York, NY~~10027
U.S.A.}

\altaffiltext{1}{
Including data from the NASA/ESA {\it Hubble Space
Telescope}, operated at the Space Telescope Science Institute, which is
operated by AURA, Inc., under NASA contract NAS 5-26555.}

\altaffiltext{2}{Visiting Astronomer, Cerro Tololo Inter-American Observatory. 
(operated by AURA, Inc., under cooperative agreement with the NSF, and Visiting
Observer, Las Campanas Observatory, of the Observatories of the Carnegie
Institution of Washington} 

\begin{abstract}
The volume around the SN~1987A contains a variety of structures,
not just the three rings glowing in recombination lines.
Many of these are revealed by light echoes, so are mapped
in three dimensions by our optical imaging of the SN environs.
The rings reside in a bipolar nebula containing them at its waist
and crowns, and which is itself contained in a larger, diffuse nebula with a
detectable equatorial overdensity.
This diffuse nebula terminates in a denser wall which likely
marks the inner edge of a bubble blown by the progenitor's main sequence wind.
Along with mapping these structures, we
measure spectroscopically the velocity of the gas,
revealing, for instance, kinematic ages for the inner and outer rings
in close agreement with each other.
The presence of these structures, their ages and morphologies must be
included in models explaining the evolution of the progenitor star and its
mass loss envelope.
\end{abstract}


\section{Introduction}

The nebula around SN 1987A is complex: not only the
three rings now seen easily via recombination radiation,
but further structure seen in other ways, and
inferred from the already interacting SN ejecta.
The presence of material of the inner ring was
discovered with $IUE$ as narrow emission lines (Fransson et al.~1989), and
shown to be spatially resolved from the ground (Wampler \& Richichi 1989).
The outer rings (the northern ``NOR'' and southern ``SOR'') and the
approximate shape of the inner, equatorial ring (``ER'') were
found by Crotts et al.~(1989).
The ring structure was further resolved using the NTT
(Wampler et al.~1990) and $HST$ (Jakobsen et al.~1991, Burrows et
al.~1995).
The kinematics of the ER indicate a true ring, and not a limb-brightened
spheroid (Crotts \& Heathcote 1991).
The rings are connected by a double-lobed nebula with the ER at its waist
and likely terminating at its outer extremes at the ORs (Crotts et al.~1995).
A light echo just outside the rings (Bond et al.~1991),
encompasses a diffuse medium of echoing material (Crotts and Kunkel 1991).
This diffuse medium includes an equatorial overdensity in the
same plane as the ER (Crotts et al.~1995).
Here we present further, unpublished results bearing on the
progenitor star's nature and the production of this circumstellar nebula.

\section{Kinematics of the Three Rings}

The velocity field in [N~II]$\lambda$6583 emission shows a gradient across the
ER minor axis like that of a ring, inclined at $43^\circ$, expanding radially
at $v_{exp} = 10.3$~km~s$^{-1}$ (Crotts \& Heathcote 1991).
Other groups find 10.3, 8.3 and 11~km~s$^{-1}$ (Cumming 1994, Meaburn et
al.~1995, Panagia et al.~1996, respectively), confirmed
(10.5$\pm0.3$~km~s$^{-1}$) with $HST$ STIS spectroscopy (Crotts \& Heathcote
2000).

More novel is combining the angular resolution of $HST$ and spectral resolution
of CTIO 4m echelle spectra ($\sim8$~km~s$^{-1}$ FWHM) to separate ER and OR
signals (Crotts \& Heathcote 2000).
By chance, OR segments more than 1~arcsec from the SN have velocities close to
the ER's, while those near the ER differ from it by up to 30 km s$^{-1}$.
Echelle spectra or $\sim100$~km~s$^{-1}$-resolution STIS
data together dissect both regimes of the nebula.
For signals distinct from the ER, spatial position indicates whether
they arise from the SOR or NOR.
We find NOR velocities, relative to the 289~km~s$^{-1}$ of the SN, of $+1$ to
$+24$~km~s$^{-1}$, and $+3$ to $-23$~km~s$^{-1}$ for the SOR.
The NOR is geometrically similar to the ER, while the SOR is rounder but ovoid.
For the SOR we use an inclination alternatively the same as the NOR's or that
which makes it nearly round when de-projected, $43^\circ$ and $31^\circ$,
respectively.
The kinematic ages implied by these velocities, assuming homologous expansion,
are 21700y for the NOR, and 19900y or 20800y for the SOR.
A similar ratio for the ER yields 19500y (Crotts \& Heathcote 1991), consistent
with the ORs (with the $1\sigma$ errors of $\approx 1500$y).
Full interpretation of these results await successful modeling of pre-SN
progenitor evolution, but these ages run counter to statements concerning $HST$
FOS spectra (Panagia et al.~1996) that the NOR is nitrogen-poor versus the
ER, and that this underabundance results from the ORs being ejected $\sim
10^4$y before the ER.

\section{Echoing Circumstellar Matter Beyond the Rings}

Echo mapping producing three-dimensional maps (Crotts et al.~1995), applied to
the region just outside the rings, reveal four features:
1) an oval of 9-15 arcsec radius (depending on observation epoch),
modeled by Chevalier and Emmering (1989) as a
contact discontinuity (CD) between the red supergiant (RSG) wind and the
surrounding bubble blown by the main-sequence (MS) progenitor wind.
2) a sheet of material along the equatorial plane defined by the ER and
bisecting the bipolar nebula (c.f.~Crotts et al.~1995, Fig.~21);~
3) a diffuse echo (Crotts \& Kunkel 1991) extending from the bipolar nebula
outward to feature \#1, and~
4) ``Napoleon's Hat,'' (Wampler et al.~1990) apparently a
discontinuity in the gradient of feature \#3.
We followed these since 1988-9 until their disappearance, and highlight
feature \#1 here.
Now we are pursuing these features to fainter surface brightness
using improved image subtraction e.g.~Crotts \& Tomaney (1996).

This diffuse medium indicates the duration of RSG mass loss.
If the velocities seen at the inner edge of this wind e.g.~at the ER, NOR and
SOR apply throughout, $v_{exp} \approx 20$~km~s$^{-1}$, the age of this diffuse
structure is then $\ga 2\times 10^5$y.

% figure 1
\begin{figure}
\vspace{2.35in}
%\vspace{2.85in}
\plotfiddle{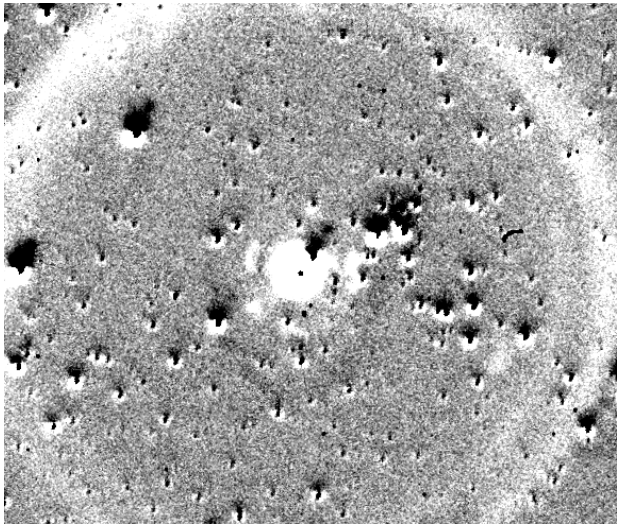}{-1in}{0}{100}{100}{-280}{-890}
\caption{
An 85-arcsec wide residual image showing, at increasing radii, echoes from the
diffuse circumstellar nebula, contact discontinuity and interstellar medium
(130~pc from the SN).
With permanent nebulosity and stellar flux removed (but some
stellar residuals remaining), the echoes become more visible.
Contained within the diffuse echo are the bipolar nebula and rings.
This image was obtained on day 750 at the LCO 2.5-meter by W.~Kunkel, in a
band centered at 6023\AA.
} \label{fig-1}
\end{figure}

The CD echo lies in fragmented ovals and faded drastically in 1991-2.
We centroid on individual echo patches in $(x,y)$ and calculate line-of-sight
depth $z=(x^2+y^2)/2ct-ct/2$, where $t$ is time since maximum light.
Figure 2 shows two views of the CD (and the bipolar nebula near the origin) as
seen from vantage points perpendicular to the sightline to Earth.

% figure 2
\begin{figure}
\vspace{2.00in}
%\vspace{2.25in}
\plotfiddle{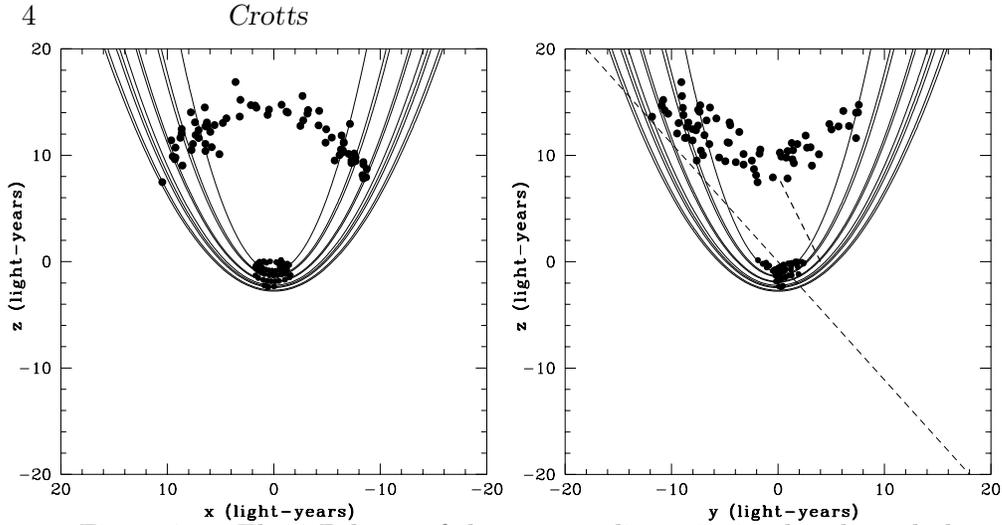}{0in}{0}{100}{100}{-225}{-64}
\caption{The 3-D locus of the contact discontinuity,
bipolar nebula and volume probed by light echoes.~~
$Left$: the view from far to the north ($x$ is distance in right ascension).
The CD wraps around the SN at radius $\sim$4~pc.
Points sampled must lie within the light-echo parabolae
for the epochs of observation.~~~
$Right$: view of the CD and bipolar echo seen
from far to the east ($y$ is distance in declination).
The thick dashed line extending from the northern end of the bipolar nebula
to the CD at the same $y$ as the SN denotes Napoleon's Hat.
The diagonal, thin dashed line is the symmetry axis of the bipolar nebula.
} \label{fig-2}
\end{figure}

\begin{figure}
\plotfiddle{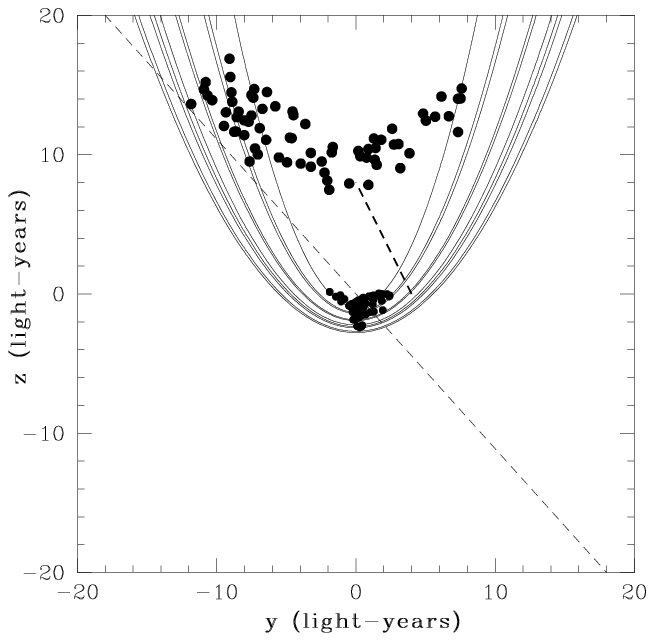}{0in}{0}{100}{100}{-35}{082}
\end{figure}

If the ``CD'' is a true contact discontinuity (Chevalier \& Emmering 1989)
between the RSG wind and the bubble blown by the MS blue supergiant progenitor,
the pressure in this bubble should be nearly uniform over a few parsecs, with
the shape of the CD due primarily to ram pressure $\rho v^2$ against nearly
uniform bubble pressure.
If the CD is aspherical, this is due to anisotropic $\rho v^2$.
Given the equatorial overdensity (feature \# 2 above) in the RSG wind, a polar
bulge or even a spherical CD indicates a RSG velocity at the poles higher than
at the equator.
This wind configuration has not been considered in interacting wind models
e.g.~Blondin \& Lundqvist (1993), Martin \& Arnett (1995), Blondin et
al.~(1996).
The presence of the CD and the diffuse material inside it runs counter to some,
more complicated geometries proposed for the circumstellar environment of
SN~1987A (c.f.~Podsiadlowski et al.~1991); indeed we know of no model that
successfully incorporates it and the three rings.

{\bf Acknowledgements.}
This work reflects major efforts by Bill Kunkel and Steve
Heathcote, and improvements underway by Ben Sugerman.
We appreciate the continued encouragement of Cerro Tololo and Las Campanas
observatories.

\end{document}